\documentclass[10pt,twoside]{article}
\usepackage{graphicx}
\usepackage{psfig}
\makeindex

\begin{document}

\title{A Simplified Model \\
of the Formation of Structures in Dark Matter}

\markboth{O.Yu.\,Tsupko and G.S.\,Bisnovatyi-Kogan}{Formation of
Structures in Dark Matter}

\author{O.Yu.\,Tsupko$^{1,3}$, G.S.\,Bisnovatyi-Kogan$^{1,2}$
\\[5mm]
\it $^1$Space Research Institute of Russian Academy of Science, Moscow, Russia\\
\it $^2$Joint Institute Nuclear Research, Dubna, Russia\\
\it $^3$Moscow Engineering Physics Institute, Russia\\
\it e-mail: gkogan@mx.iki.rssi.ru, tsupko@iki.rssi.ru\\
}

\date{}
\maketitle

\thispagestyle{empty}

\begin{abstract}
\noindent The collapse of a rotating 3-axis ellipsoid is
approximated by a system of ordinary differential equations.
Violent relaxation, mass and angular momentum losses are
taken into account phenomenologically. The formation of the
equilibrium configuration and different types of instability
are investigated.\\

\end{abstract}

According to modern cosmological ideas the most part of matter in
the Universe belongs to so called Cold Dark Matter, consisting of
non-relativistic particles.

The study of the formation of dark matter objects in the Universe
is based on N-body si\-mulations, which are very time consuming.
In this situation a simplified approach may become useful.

Here we derive equations for the dynamical behaviour of a
compressible rotating 3-axis ellipsoid in which a motion along
axes takes place in the common gravitational field of a uniform
ellipsoid and under the action of the isotropic pressure. Collapse
in the dark matter are characterized by non-collisional
relaxation, based on the idea of a "violent relaxation" of
Lynden-Bell \cite{ref2}. So the collapse of the ellipsoid is
approximated by a system of ordinary differential equations, where
the relaxation and losses of energy, mass and angular momentum are
taken into account phenomenologically. The system is solved
numerically.

Bisnovatyi-Kogan\cite{ref1} considered the case of spheroid ($a =
b \neq c$), where there are analytical formulaes for the
gravitational potential and forces.

Let us consider a uniform 3-axis ellipsoid, consisted of
non-collisi\-onal non-relativistic particles, with semi-axes $a
\neq b \neq c$ and rotating uniformly with an angular velocity
$\Omega$ around the axis $z$.

The mass $m$ and total angular momentum $M$ of a uniform ellipsoid
are connected with uniform density, angular velocity and semi-axes
as $ \label{mass-moment}m = \frac{4\pi}{3} \, \rho \, abc \, ,
\quad M = \frac{m}{5} \, \Omega (a^2+b^2) \, . $ Assume a linear
dependence of the velocity on the coordinates: $
\label{linear-dep} \upsilon_x =\dot{a}x/a \: $, \, $\upsilon_y
=\dot{b}y/b \: $, \, $ \upsilon_z =\dot{c}z/c \, . $ The
gravitational energy of the uniform ellipsoid is defined as:
$$
\label{grav-energy}
U_g=-\frac{3Gm^2}{10}\int\limits_0^{\infty}\frac{du}{\sqrt{(a^2+u)(b^2+u)(c^2+u)}}
$$
and is expressed in elliptical integrals.

Consider a compressible ellipsoid with a constant mass and angular
momentum, a total thermal energy of non-relativistic dark matter
particles $E_{th} \sim V^{-2/3} \sim (abc)^{-2/3}$ and the
relation between pressure $P$ and thermal energy $E_{th}$ as
$E_{th} = \frac{3}{2} \frac{P}{V}$. In absence of any dissipation
the ellipsoid is a conservative system.

To derive equations of motion let write for it a Lagrange function
$$
\label{Lagr-funct} L = U_{kin} - U_{pot}\, , \: U_{pot} = U_g +
E_{th} + U_{rot} \: , \quad U_{kin} = \frac{m}{10} \,
(\dot{a}^2+\dot{b}^2+\dot{c}^2) \: ,
$$
$$
\label{E-th} E_{th} = \frac{E_{th,in} (a_{in}b_{in}c_{in})^{2/3}}
{(abc)^{2/3}} = \frac {\varepsilon} {(abc)^{2/3}} \: , \: \:
U_{rot} = \frac{1}{2}\,\rho\int \limits_V V_{rot}^2 \,dV =
\frac{5}{2} \, \frac{M^2}{m(a^2+b^2)} \: .
$$
By variation of the Lagrange function we obtain Lagrange equations
of motion. It is easy to check that equilibrium solution of these
equations is the Maclaurin sphe\-roid, and in addition the 3-axis
Jacobi ellipsoid appears at large angular momentum.

In reality there is the "violent relaxa\-tion" in the
collisionless system \cite{ref2}. Therefore there is a drag force,
which is described phenomenologically by adding of the terms $ -
\frac{\dot{a}}{\tau_{rel}} \: , \quad - \frac{\dot{b}}{\tau_{rel}}
\: , \quad - \frac{\dot{c}}{\tau_{rel}} $ in the right-hand parts
of equations of motions.

Here we have scaled the relaxation time $\tau_{rel}$ by the Jeans
characteristic time with a constant value of $\alpha_{rel}$ : $
\label{tau-rel} \tau_{rel} = \alpha_{rel} \tau_J = 2 \pi \,
\alpha_{rel} \sqrt{\frac{abc}{3Gm}} $ .

The process of relaxation is accompanied also by energy, mass and
angular momentum losses from the system. These losses may be
described phenomenologically by characteristic times $\tau_{el},
\tau_{ml}, \tau_{Ml}$.

To obtain a numerical solution of equations we write them using
non-dimensional variables. Taking into account mass and angular
 momentum losses, the dynamics of the system is described by the
 following non-dimensional system of equations
$$
\ddot{a} = - \frac{\dot{a}}{m} \frac{dm}{dt} - \frac{3m}{2} \, a
\int\limits_0^{\infty}\frac{du}{(a^2+u)\sqrt{(a^2+u)(b^2+u)(c^2+u)}}
\,\, + \,\frac{10}{3m}\, \frac{1}{a} \,\frac {\varepsilon}
{(abc)^{2/3}} \,\,\,+
$$
$$+ \frac{25 M^2}{m^2} \, \frac{a}{(a^2+b^2)^2} \,\,\,
- \,\frac{\dot{a}}{\tau_{rel}} \, , $$
$$
\ddot{b} = - \frac{\dot{b}}{m} \frac{dm}{dt} - \frac{3m}{2} \, b
\int\limits_0^{\infty}\frac{du}{(b^2+u)\sqrt{(a^2+u)(b^2+u)(c^2+u)}}
\,\,+ \,\frac{10}{3m}\, \frac{1}{b} \,\frac {\varepsilon}
{(abc)^{2/3}} \,\,\,+
$$
$$+ \frac{25 M^2}{m^2} \, \frac{b}{(a^2+b^2)^2}
- \,\frac{\dot{b}}{\tau_{rel}} \, , $$
$$
\ddot{c} = - \frac{\dot{c}}{m} \frac{dm}{dt}
 - \frac{3m}{2} \, c
\int\limits_0^{\infty}\frac{du}{(c^2+u)\sqrt{(a^2+u)(b^2+u)(c^2+u)}}
\,\,+ \,\frac{10}{3m}\, \frac{1}{c} \,\frac {\varepsilon}
{(abc)^{2/3}} - \,\frac{\dot{c}}{\tau_{rel}},
$$
$$
\dot{\varepsilon} = (abc)^{2/3}\, U_{kin} \left[\left(
\frac{2}{\tau_{rel}} - \frac{1}{\tau_{el}} - \frac{2}{\tau_{ml}}
\right) - \frac{U_{rot}}{U_g} \left(\frac{2}{\tau_{Ml}} -
\frac{1}{\tau_{ml}} \right) + \frac{U_{kin}}{U_g \,
\tau_{ml}}\right] ,
$$
$$
\dot{m} = - \frac{1}{3\tau_{ml}} \,
(\dot{a}^2+\dot{b}^2+\dot{c}^2)
\left[\int\limits_0^{\infty}\frac{du}{\sqrt{(a^2+u)(b^2+u)(c^2+u)}}\right]^{-1}
, \frac{M}{M_{in}} =
\left(\frac{m}{m_{in}}\right)^{\frac{\tau_{ml}}{\tau_{Ml}}} .
$$

This system is solved numerically for several initial parameters.

The case of spheroid ($a = b \neq c$) is consi\-dered by
Bisnovatyi-Kogan \cite{ref1}. Our 3-d equations give the same
result for spheroidal initial conditions.

In the case of 3-axis ellipsoid there are new qualitative effects,
because there is additional degree of freedom compared to the case
of spheroid. At large initial angular momentum $M_{in}$ we observe
the instability of the Mac\-laurin sphe\-roid for the
tra\-ns\-for\-ma\-tion into the Jacobi el\-lip\-soid. So in this
case the system rea\-ches equilibrium 3-axis ellipsoid. See Fig.1.

\begin{figure}[h]
\centerline{\hbox{\includegraphics[width=0.6\textwidth]{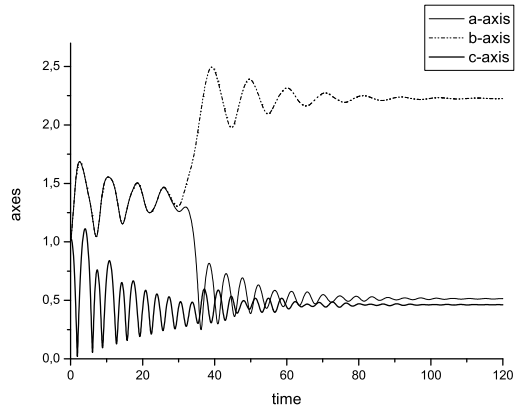}}}
\caption{The instability at large momentum.}
\end{figure}

At small initial angular momentum $M_{in}$ the system rea\-ches
the oblate spheroid. But we have an instability, disappeared
during the relaxation. It is the instability of system with purely
radial trajectories.

The work is in progress. We are going to investigate the different
aspects of instability in more detail.

The authors are thankful to the RFRF grant No. 02-02-16900 for
partial support.

\end{document}